\documentclass[universe,review,accept,oneauthor,pdflatex]{Definitions/mdpi} 
\firstpage{1} 
\pubvolume{1}
\issuenum{1}
\articlenumber{0}
\pubyear{2022}
\copyrightyear{2022}
\externaleditor{{Academic Editor: Sergei B. Popov, Ziri Younsi and Zdenek Stuchlik}} 
\datereceived{} 
\dateaccepted{} 
\datepublished{} 
\hreflink{https://doi.org/} 

\Title{GRMHD Simulations and Modeling for Jet Formation and Acceleration Region in AGNs}

\TitleCitation{GRMHD Simulations and Modeling for Jet Formation and Acceleration Region in AGNs}

\Author{Yosuke Mizuno
 $^{1,2,3}$\orcidA{}}

\AuthorNames{Yosuke Mizuno}

\AuthorCitation{Mizuno, Y.}

\address{%
$^{1}$ \quad Tsung-Dao Lee Institute, Shanghai Jiao Tong University, Shanghai~201210, People’s Republic of China; mizuno@sjtu.edu.cn \\
$^{2}$ \quad School of Physics and Astronomy, Shanghai Jiao Tong University, Shanghai~200240, People’s Republic of China \\
$^{3}$ \quad Institut f\"{u}r Theoretische Physik, Goethe Universit\"{a}t, Max-von-Laue Str. 1, \mbox{D-60438 Frankfurt am Main, Germany}}

\abstract{Relativistic jets are collimated plasma outflows with relativistic speeds. Astrophysical objects involving relativistic jets are a system comprising a compact object such as a black hole, surrounded by rotating accretion flows, with the relativistic jets produced near the central compact object. The most accepted models explaining the origin of relativistic jets involve magnetohydrodynamic (MHD) processes. Over the past few decades, many general relativistic MHD (GRMHD) codes have been developed and applied to model relativistic jet formation in various conditions. This short review provides an overview of the recent progress of GRMHD simulations in generating relativistic jets and their modeling for observations. }


\begin{document}

\section{Introduction}

Relativistic jets are amongst the most powerful astrophysical phenomena discovered to date. Their relativistic nature causes them to emit powerful and extremely time-variable radiation in all ranges of wavelength, from radio to gamma rays. This makes them detectable at cosmological distances. Relativistic jets are known to be launched as the result of accretion processes onto extremely compact objects such as black holes (BHs) in the presence of rotating accretion flows and magnetic fields. This makes relativistic jets a powerful tool to probe the environment of objects in extremely compact matter states, and the physics of high-energy plasmas and their magnetic fields on different scales.

In the accretion processes onto BHs, a substantial fraction of the gravitational binding energy of the accreting matter is released within tens of gravitational radii from the BH. This released energy supplies the powerful radiation. Since the radiated energy originates from the vicinity of the BH, a fully general relativistic treatment is essential for the modeling of these objects and the flows of plasma in their vicinity.

Several mechanisms of jet flow acceleration and collimation have been proposed. These includes gas-pressure acceleration, acceleration by radiation, and magnetohydrodynamic (MHD) processes (e.g., \cite{Begelman1984}). It is also possible that different mechanisms operate in different sources \cite{deGouveiaDalPino2005}, or, otherwise, that different mechanisms are operating simultaneously~\cite{Beskin2010}. Currently, the most promising mechanism is that the jets may arise from the combined effects of magnetic fields and rotation. The important mechanisms here are the Blandford-Znajek (BZ) \cite{Blandford1977} and the Blandford-Payne (BP) models \cite{Blandford1982,Bisnovatyi-Kogan2001}. In BP models, the jet is formed as a result of magnetocentrifugal acceleration of matter from the surface of an accretion disk. On the other hand, in the BZ model relativistic jets can be launched from the black hole magnetosphere by extracting rotational energy of BHs. From these two models, we believe the relativistic Poynting-flux-dominated (energy and angular momentum outflow carried predominantly by the electromagnetic field) jets are driven by rotational energy of the BHs as invoked in the BZ model, whereas the sub-relativistic matter-dominated jets/winds are driven by rotational energy of accretion flow owing to a magnetocentrifugal mechanism as in the BP model. However, there can be other alternative mechanisms, such as the gradient of magnetic and gas pressure. If the jet has sufficiently large specific enthalpy and is overpressured, the relativistic jets can be powerfully boosted by the propagation of a rarefaction wave from the interface between jet and ambient medium (e.g., \cite{Aloy2006,Mizuno2008}). 

The essential physics for AGN jets can now be captured in relativistic MHD (RMHD) simulations. In particular, in order to understand jet formation from the vicinity of BHs, general relativistic MHD (GRMHD) simulations are required. Over the past few decades, many GRMHD codes have been developed (e.g., \cite{Hawley1984,Koide2000,DeVilliers2003,Gammie2003,Baiotti2005,Duez2005,Annios2005,Anton2006,Mizuno2006,DelZanna2007,Giacomazzo2007,Etienne2015,White2016,Zanotti2015,Porth2017,Olivares2019,Liska2019b,Chi-kitCheong2020}) employing the 3 + 1 decomposition of spacetime and conservative ‘Godunov’ schemes based on approximate Riemann solvers~\cite{Rezzolla2013,Font2003,Marti2015}. 
These codes are applied to study a variety of high-energy astrophysical phenomena. Some of these GRMHD codes incorporate radiation (e.g., \cite{Sadowski2013,McKinney2014,Takahashi2016}), and/or non-ideal MHD processes (e.g., \cite{Bucciantini2013,Dionysopoulou2013,Chandra2015,Chandra2017,DelZanna2018,Ripperda2019}). 
In state-of-the-art GRMHD codes, full treatment of adaptive mesh refinement has been implemented (see \cite{Porth2017,Olivares2019,White2016,Liska2019b}) which is useful for obtaining higher spatial resolution in particular interesting regions such as strong shocks, turbulence, and shear regions. 

Depending on the mass accretion rate, a black hole accretion system can be found in various spectral states \cite{Fender2004,Markoff2005}. Some AGNs have radiative power $L$ in excess of their corresponding Eddington luminosity ($L_{\rm Edd}$). At the Eddington luminosity, radiation forces balance the gravity of the central object. 
In accretion disk theory, one of the scale parameters is mass accretion rate, where the Eddington mass accretion rate is $\dot{M}_{\rm Edd} \equiv L_{\rm Edd}/\epsilon c^2$, where $\epsilon$ is radiative efficiency $\epsilon \equiv L/\dot{M}c^2$. In some supermassive BHs, including the primary targets of observations by the Event Horizon Telescope Collaboration (EHTC), i.e., Sgr A* and M87, their mass accretion rates are well below the Eddington accretion rate, $\dot{M} \ll 10^{-2} \dot{M}_{\rm Edd}$ \cite{Marrone2007,Ho2009}. In this regime, the accretion flow advects most of the viscously released energy into the BH rather than radiating it to infinity. Such optically thin, radiatively inefficient, and geometrically thick flows are so-called advection-dominated accretion flows (ADAFs, see \cite{Narayan1994,Narayan1995,Abramowicz1995,Yuan2014}). 
Analytical and semi-analytical approaches are reasonably successful in reproducing the main features in the spectra of ADAFs \mbox{(see, e.g., \cite{Yuan2003})}. However, numerical GRMHD simulations are essential to gain an understanding of the detailed physical processes.
The accreting gas in an ADAF is radiatively inefficient. Therefore, an ADAF is also referred to as a radiatively inefficient accretion flow (RIAF). 

At higher mass accretion rates, $10^{-2} \le \dot{M}/\dot{M}_{\rm Edd} \le 1$, radiative cooling becomes effective and the inner accretion disk shrinks into an optically thick geometrically thin accretion disk (e.g.,~\cite{Narayan1995}). This state is the so-called standard thin accretion disk (e.g.,~\cite{Novikov1973,Shakura1973}). In thin accretion disk theory, the disk angular momentum is transported by $\alpha$ viscosity and its spectrum is well described by thermal black-body radiation.  

At super-Eddington accretion rates, $\dot{M}/\dot{M}_{\rm Edd} \ge 1$, accretion flows again become radiatively inefficient. As the optical depth is large, the photon diffusion timescale from the disk interior to the photo-sphere becomes longer than the accretion timescale. Thus, the photons are advected with the accreting matter onto the black hole. Such a state is the so-called slim accretion disk \cite{Abramowicz1988}. Clearly, in order to understand accretion physics in these systems, radiation MHD models with self-consistently coupled gas, radiation, and magnetic fields are important. 

In this short review, I will overview the recent progress in the study of relativistic jet formation via GRMHD simulations and the modeling of relativistic jets based on the results of GRMHD simulations. In Section \ref{sec2}, I discuss jet formation from geometrically thin accretion disks. Jet formation from geometrically thick accretion torii is discussed in Section \ref{sec3}. I then present an overview of radiative GRMHD simulations and the effects on jet formation in Section \ref{sec4}. I briefly touch on some GRMHD simulations for jet formation without accretion disks in Section \ref{sec5}. Then, in Section \ref{sec6}, I discuss jet modeling work using the results of GRMHD simulations. Finally, I summarize the current progress and findings in Section \ref{sec7}.   

\section{Jet Formation from Geometrically Thin Accretion Disk}\label{sec2}



Pioneering work of relativistic jet formation via GRMHD simulations was performed by \citet{Koide1998,Koide1999} in the consideration of a Keplerian thin accretion disk with a strong vertical magnetic field around a non-rotating black hole. In these two-dimensional (2D) simulations, the matter in the disk loses angular momentum by magnetic breaking and then falls into the black hole. A centrifugal barrier near the black hole horizon decelerates the infalling material and produces a standing shock. Plasma near the shock is accelerated by the Lorentz force and forms bipolar jets. Inside these magnetically driven jets, the gradient of gas pressure also generates a jet above the shock region (gas-pressure-driven jets). Such two-layered jets are formed in both a hydrostatic corona and a radially infalling corona. \citet{Koide2000} extended this work to rotating black holes and found that similar two-layered jets are produced in both co-rotating and counter-rotating black hole magnetospheres. This simulation was extended to three dimensions (3D) by \citet{Nishikawa2005}. 

After this, \citet{Hardee2007} performed similar GRMHD simulations of Keplerian thin disks with vertical magnetic fields, now considering different black hole spins. In this study, the formation of two-component jets due to black hole spin effect is demonstrated.
 The resulting jet is mildly relativistic ($\sim$$0.5c$). Due to numerical limitations, these earlier simulations did not investigate the long-term evolution. 

Thin disk simulations have been used for the investigation of the validity of Novikov and Thorne disk theory \cite{Novikov1973}, finding around a 10\% deviation of radiative efficiency (e.g.,~\mbox{\cite{Shafee2008,Noble2010,Noble2011,Penna2010}}). In much stronger magnetic field cases, such as magnetically arrested disk states, such deviations becomes much larger, reaching up to $\sim$$80\%$ \cite{Avara2016}.  
Recently, \citet{Dihingia2021} has shown the formation of a BZ-jet from a black hole and BP disk wind from a thin accretion disk with an inclined poloidal magnetic field geometry (see Figure~\ref{fig:thin_disk_jet}). These authors also found that plasmoids are formed in thin accretion disks due to magnetic reconnection. Near the polar axis, there are strong magnetized regions in which the BZ process operates and can launch jets. The jet velocity can reach around $\gamma \sim 10$. This result indicates that strong initial poloidal magnetic fields are essential for creating strong jet and disk winds from thin accretion disks.  

\begin{figure}[H]	
\includegraphics[width=0.69\textwidth]{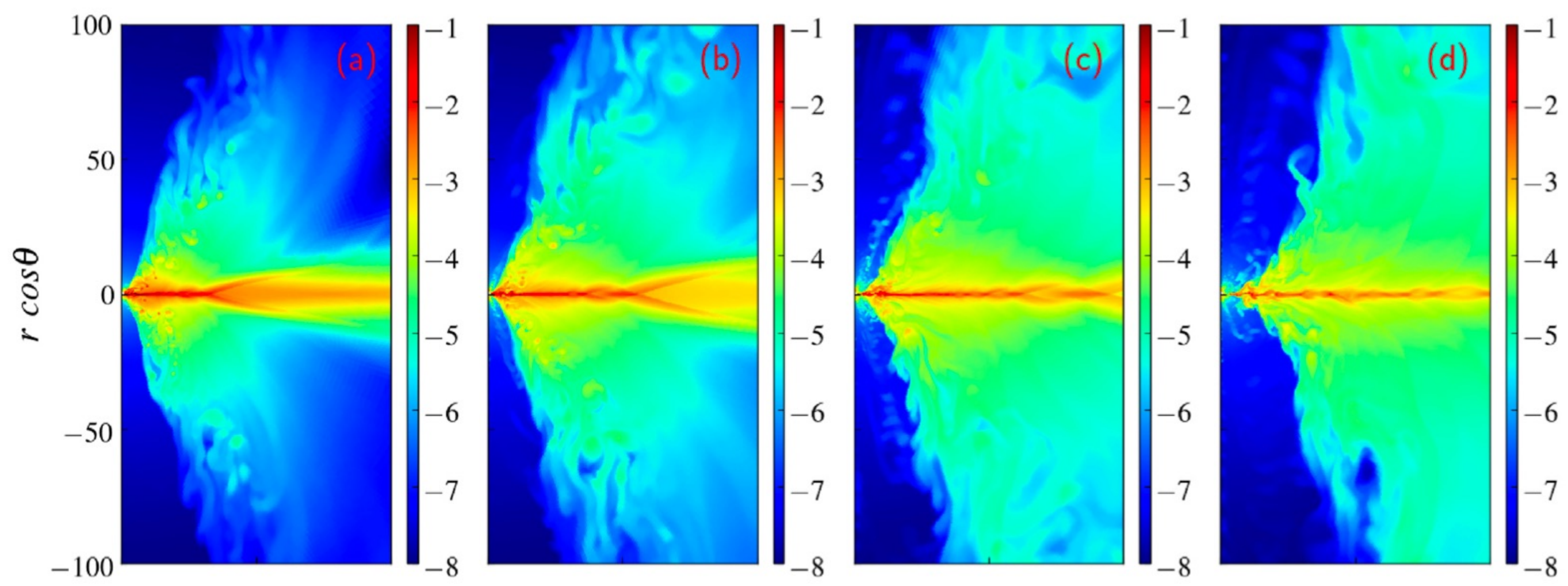}
\caption{Evolution of 2D GRMHD simulations of a geometrically thin disk around a rotating black hole at $t=500$ ({\bf a}), $t=1000$ ({\bf b}), $t=2000$ ({\bf c}), and $t=4000$ ({\bf d}). The color contour is logarithmic normalized density ($\rho/\rho_{\rm max}$). Figure is reproduced with permission from Dihingia et al. MNRAS, 505, 3596 (2021) \cite{Dihingia2021}.}
\end{figure}  

\subsection{Jet Formation from Tilted Thin Disks}

\begin{figure}[H]	
\includegraphics[width=0.4\textwidth]{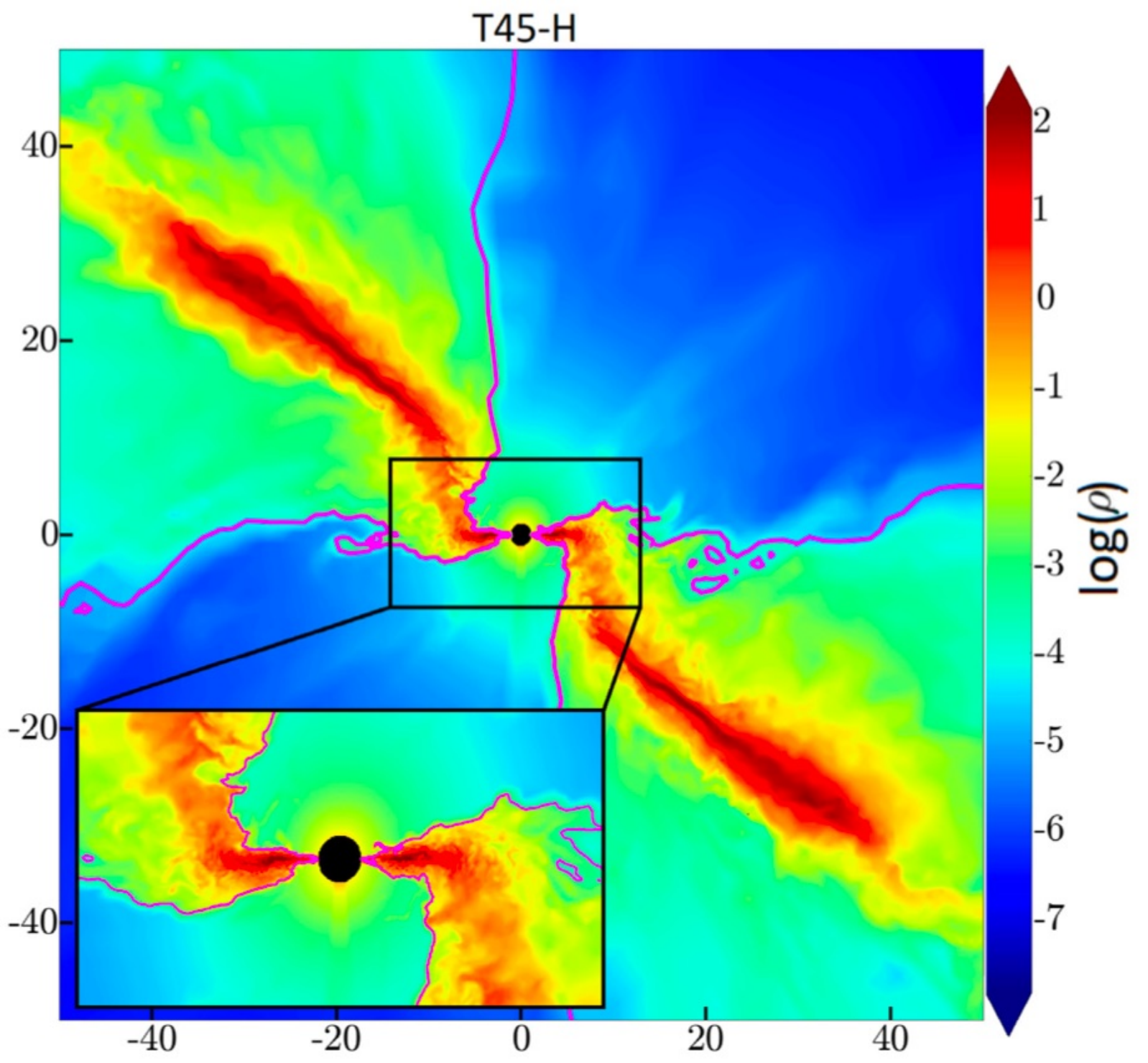}
\caption{Snapshot of a vertical slice from a 3D GRMHD simulation of a tilted thin disk. The color contour shows logarithmic density. Magenta lines indicate the jet boundary, defined as $p_b = 5\rho c^2$. The results present the alignment of the inner disk along the equator. Figure is reproduced with permission from Liska et al. MNRAS, 507, 983 (2021) \cite{Liska2021}}.  \label{fig:tilted_thin_jet}
\end{figure}

The misalignment between the accretion disk and the black hole angular momentum vector is another degree of freedom of GRMHD simulations. Important changes in the dynamics of an accreting system exhibiting such misalignment are thought to be a result of Lense--Thirring (LT; Lense and Thirring, 1918) precession. LT precession is a general relativistic frame dragging effect where test particles on tilted orbits around the central object precess with a radially dependent angular frequency $\Omega_{\rm LT} \propto 1/r^3$. \citet{Bardeen1975} showed that a viscous disk would be expected to relax to a configuration where the inner region becomes aligned with the equatorial plane of the black hole. \citet{Liska2018} have investigated the Bardeen--Petterson alignment using high-resolution GRMHD simulations with thin accretion disks, showing the alignment in the inner region of the disk (see Figure~\ref{fig:tilted_thin_jet}). They have also demonstrated that powerful relativistic jets are still launched in this misaligned case. Interestingly, in the case of highly tilted disks, the disk is torn apart and forms a rapidly precessing inner sub-disk with a slowly precessing outer sub-disk \cite{Liska2021}. The resulting jet precesses rapidly together with the inner sub-disk.

\subsection{Thin Disk Simulations in Non-Ideal GRMHD}

GRMHD simulations of jet launching from thin accretion disks have been extended to the non-ideal regime by including resistivity \cite{Qian2018,Vourellis2019}. It was found that magnetic diffusivity lowers the efficiency of accretion and ejection. The launched jet and wind become weaker and slower ($\simeq$$0.1c$).  

In general, radiation effects are an important property in modeling thin accretion disks, in particular, radiative cooling. However, most GRMHD simulations of thin accretion disks do not include such effects due to the numerical difficulty and computational cost. Therefore, in order to obtain a more realistic picture of jet formation from thin accretion disks, radiation GRMHD simulations need to be performed.

\section{Jet Formation from Geometrically Thick Magnetized Torii}\label{sec3}

Modern BH accretion disk theory suggests that angular momentum transport is provided by Maxwell and Reynolds stresses within the orbiting plasma. MHD turbulence is driven by the magnetorotational instability (MRI) within a differentially rotating disk~\mbox{\cite{Bulbus1991,Bulbus1998}}. Convective motions developed by MRI are a general phenomenon in RIAFs. Since the viscosity that drives accretion originates from MRI, magnetic fields play a crucial role. 
In order to develop MRI in the accretion disk, in general, a geometrically thick accretion torus with a weak poloidal magnetic loop inside the torus is considered. The disk thickness ($H/r$, where $H$ is height of the disk and $r$ is the radius from a black hole) is a key parameter for capturing the MRI in numerical simulations. If $H/r$ becomes smaller, much greater numerical resolution is needed to resolve the accretion disk. Therefore, geometrically thick torii are favored as the initial setup for numerical simulations.

\begin{figure}[H]	
\includegraphics[width=0.69\textwidth]{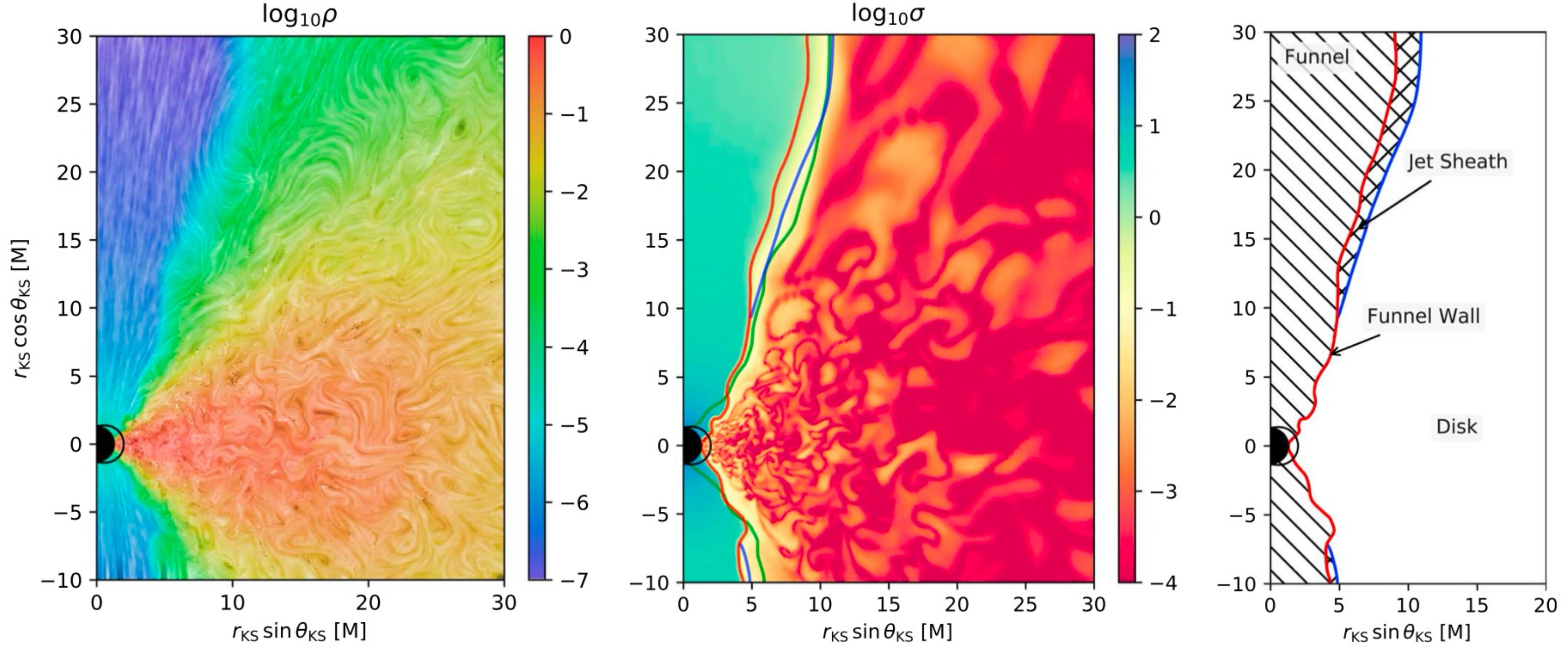}
\caption{Snapshot of GRMHD simulations of geometrically a thick torus in the SANE regime. Left: color contour shows logarithmic density and white lines indicate rendering of the magnetic field structure using line-integral convolution. Center: color contour shows logarithmic magnetization. The magnetized funnel is represented by $\sigma=1$ (red lines), the disk is indicated by $\beta=1$ (green lines), and the geometric Bernoulli criterion ($u_t = -1$) is given by the blue solid line in the region outside of the funnel. Right: schematic of the main components of SANE regime. Figure is reproduced with permission from Porth et al. ApJS, 243, 26 (2019) \cite{Porth2019}.}
 \label{fig:SANE_jet}

\end{figure}

Pioneering GRMHD simulations of thick accretion torii in RIAFs were performed by \citet{DeVilliers2003a} and \citet{McKinney2004}. \mbox{\citet{DeVilliers2003a}}~studied the accretion process in rotating black hole spacetimes and investigated the dependency of the accretion on the black hole spin parameter. \citet{McKinney2004} estimated the outward energy flux from the Kerr black hole horizon via the BZ process. In their simulations, a flow structure can be decomposed into a disk, a corona, a disk wind, and a highly magnetized polar funnel (see Figure~\ref{fig:SANE_jet}) (e.g., \cite{McKinney2006,McKinney2009}). Such an accretion flow regime is termed standard accretion and normal evolution (SANE). 
RIAFs with poloidal magnetic flux and a spinning black hole are key ingredients for producing Poynting-flux-dominated (PFD) funnel jets (e.g., \cite{DeVilliers2003b,DeVilliers2005,Gammie2003,Hirose2004,McKinney2004,Hawley2006,McKinney2006,Beckwith2008}). Between the PFD polar funnel jet and the disk wind, there is the region of unbound mass flux referred to as the “funnel-wall” jet. The boundary between the low-density PFD funnel jet and the high-density funnel wall jet is sharp and clear. The properties of the disk wind have been investigated for different black hole spins and magnetic field configurations (e.g., \cite{Narayan2012,Sadowski2013b}). 

From theoretical work on jets (outflows), jet structure is understood to depend on the interaction of outflows with the ambient medium. Understanding the coupling between outflows and ambient medium requires information about the magnetic field configuration, physical conditions at the jet base, and the properties of the ambient medium, because these are related to the jet's final energy contents and Lorentz factor. Many idealized semi-analytic studies have been performed to investigate jet properties (e.g., \cite{Vlahakis2003,Beskin2006,Lyubarsky2009,Pu2015}).

Comparison of GRMHD simulations with steady solutions of force-free jets have shown good agreement of these properties with the PFD funnel jet (e.g., \cite{McKinney2007a,Nakamura2018}). PFD jet profiles in time-averaged GRMHD simulations exhibit a power-law profile to the parabolic solution. This is similar to the solution of \citet{Blandford1982}. \mbox{\citet{McKinney2007b}} confirmed that similar profiles are obtained from general relativistic force-free electro-dynamics simulations of the disk wind. 

However, there semi-analytic models and GRMHD simulations disagree in the characteristics of jet acceleration. Semi-analytic models (e.g., \cite{Beskin2006}) have reported efficient acceleration of jets to nearly the maximum Lorentz factor by converting the jet's entire energy budget. Such results are confirmed by global SRMHD simulations of jets injected from a disk with the ambient medium modeled by placing a conducting wall at the jet outer boundary (e.g., \cite{Komissarov2007,Komissarov2009,Tchekovskoy2008,Tchekhovskoy2010}), which shows bulk acceleration up to $\gamma_{\infty} \sim$ 10--1000. However, such highly efficient energy conversion is not seen in GRMHD simulations \mbox{(e.g., \cite{McKinney2006})}. In \citet{McKinney2006}, the simulations are extended up to $r=10^4$ M and show $\gamma_\infty \le 10$. Acceleration is saturated beyond a few times $100$ M. Recently, \citet{Chatterjee2019} has extended the investigation of jet acceleration up to $10^5$ M in axisymmetric (i.e., 2D) GRMHD simulations. They found the development of oscillations by the interaction between jet and wind. Such oscillations drive pinch instabilities at the jet outer boundary when the jet becomes superfast (see \cite{McKinney2006}). These pinch instabilities result in a heating up of the jet by magnetic reconnection and mass loading of the jets which affects the jet deceleration. Such pinch modes may excite kink instabilities in 3D configurations. 

When jets are magnetically dominated, they are likely to experience current-driven kink instabilities, which will lead to magnetic reconnection (e.g., \cite{Begelman1998,Giannios2006,Mizuno2009,Mizuno2011,Mizuno2012,Mizuno2014,Singh2016}). A kink instability excites large-scale helical motions that may disrupt the regular structure of the magnetic field, liberating magnetic energy and potentially resulting in flaring activity like that observed from blazars (e.g., \cite{Kadowaki2021}). Such magnetic reconnection driven by kink instability-induced turbulence may be a possible mechanism for rapid magnetic dissipation of relativistic jets and high-energy particle acceleration (e.g., \cite{Medina-Torrejon2021}).

Global SRMHD simulations of jet injection and propagation in the ambient medium have shown bulk jet acceleration to occur (e.g., \cite{Komissarov2009,Tchekhovskoy2010}). When such external pressure support drops and the jet enters the the regime of ballistic expansion, additional acceleration occurs via magnetosonic rarefaction waves from the boundary between the jet and the ambient medium. This is the so-called rarefaction acceleration, which induces a conversion of magnetic energy into kinetic energy of the bulk motion (e.g., \cite{Aloy2006,Mizuno2008,Tchekhovskoy2009,Komissarov2010}). Such wave propagation in relativistic jets produces multiple chains of expansion and recollimation, with shocks and rarefaction waves (e.g., \cite{Mizuno2015}). Standing recollimation shocks in relativistic jets are related to stationary features observed in relativistic jets \cite{Gomez2016}. 

Recently, GRMHD simulations of magnetized accretion torii around black holes in an alternative theory of gravity (dilaton black hole in Einstein--Maxwell dilaton--axion theory of gravity), and even exotic compact objects like a boson star, have been performed~\cite{Mizuno2018,Olivares2020}. Although they have currently applied only to spherically symmetric black holes and exotic compact objects, these simulations are useful for understand the dynamics and testing theories of gravity from the accreting matter, as well as understanding jet formation in the different theories of gravity.

GRMHD simulations of magnetized accretion torii in the SANE regime are a standard model for GRMHD simulations. \citet{Porth2019} performed a code comparison of many existing GRMHD codes, using the same initial setups in the SANE state. In this code comparison challenge, nine GRMHD codes participated, with the results showing that agreement between GRMHD codes improves as resolution increases, obtaining consistent~results. 

\subsection{GRMHD Simulations in the MAD Regime}

\begin{figure}[H]	
\includegraphics[width=0.69\textwidth]{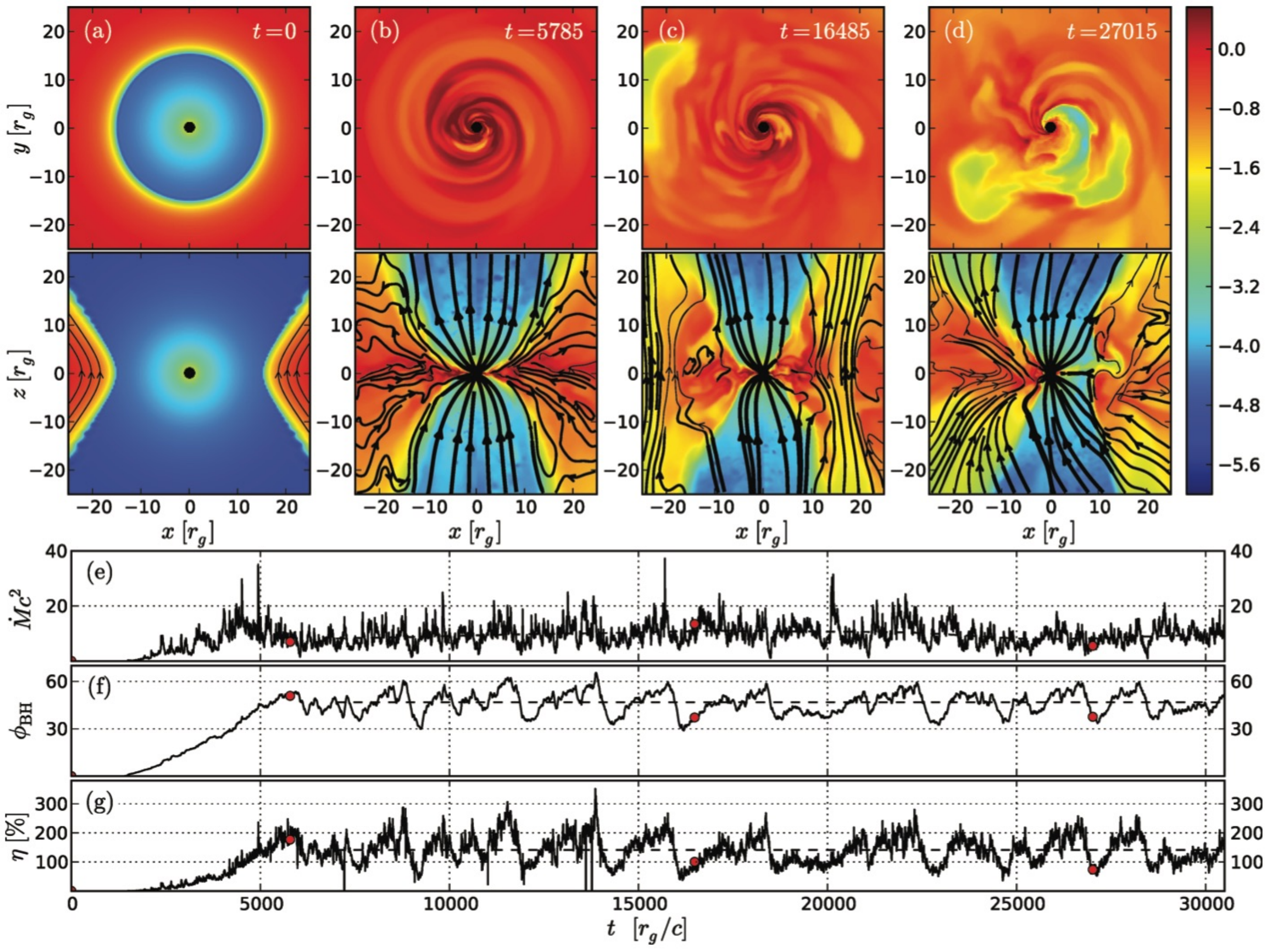}
\caption{Snapshot of
 GRMHD simulations of a geometrically thick torus in the MAD regime (Panels (\textbf{a}--\textbf{d})). The top and bottom rows show equatorial ($z = 0$) and vertical ($y = 0$) snapshots of the simulation. Color represents logarithmic density and black lines indicate magnetic field lines. Panels (\textbf{e}--\textbf{g}) show the time evolution of the mass accretion rate, magnetic flux threading the BH horizon, and the energy outflow efficiency. Figure is reproduced with permission from Tchekhovskoy et al. MNRAS, 418, L79 (2011) \cite{Tchekhovskoy2011}.} 
\label{fig:MAD_jet}
\end{figure}

One of the key properties of an accretion flow is the magnetic field, and it is particularly important to understand the accretion flow behavior
in the magnetically dominated regime. \citet{Igumenshchev2003} found that magnetic fields can become dynamically important in black hole accretion flows, preventing the inward motion of the accretion flow via magnetic pressure near the horizon (see Figure~\ref{fig:MAD_jet}) (e.g., \cite{Igumenshchev2008}). This regime is called the magnetically arrested disk (MAD) (e.g., \cite{Narayan2003}). Hot accretion flows in the MAD regime can produce powerful jets (e.g., \cite{Tchekhovskoy2011,Tchekhovskoy2012,McKinney2012}).
\citet{White2019a} have studied the effects of numerical resolution on dynamical properties in the MAD state and demonstrated convergence. 
Recently, \citet{Narayan2021} investigated black hole spin dependency in the MAD state by considering the long-term evolution of GRMHD simulations. They found that the saturated magnetic flux level and jet power in the MAD regime depend strongly on the black hole spin. The prograde spin case saturates at much higher relative magnetic flux and has more powerful jets than in retrograde cases. Similar trends are seen in \citet{Tchekhovskoy2012}. MAD simulations with spinning black holes have launched powerful jets with generalized parabolic profiles which follow the width $w \propto z^{k}$, where $z$ is jet height and power law index $k \sim$ 0.27--0.42 from the range of height $z$ = 5--100\,$r_g$, and $r_g \equiv GM/c^2$ is gravitational radius. This is similar to jets in the SANE case. The jet width also depends on the black hole spin. Prograde cases show wider jets compared to retrograde cases. Even in jet images at mm wavelengths, similar results in jet width are seen \cite{Fromm2021}. 

GRMHD simulations in the MAD regime exhibit violent episodes of flux escape from the black hole magnetosphere (e.g., \cite{Dexter2020,Porth2021}). These magnetic flux eruptions could explain the flare events observed in Sgr A* because they are associated with magnetic reconnection, which provides particle heating and acceleration during the flare events and contains enough energy to power flares.
\citet{Wong2021} have investigated the jet--disk boundary layer in different black hole spins and different accretion states, including SANE and MAD. They have shown that in the retrograde case, due to strong shear, the jet--disk boundary is unstable. This mixing layer episodically loads matter onto trapped field lines where it is forced to co-rotate with the BH, and move outward into the jet.

%

Recent polarized observations of M87 by the Event Horizon Telescope indicate that the accretion flows in M87 will be in the MAD state \cite{EHT2021b}. Other discussions of observed jet power in AGNs also suggest that MAD is favored in those systems \cite{Zamaninasab2014,Nemmen2015}. 

\subsection{Different Magnetic Field Configurations}

Magnetic field configuration is another important ingredient of accretion flows which result in jet production. In standard GRMHD simulations of magnetized accretion torii, single poloidal magnetic loops inside the torus are considered and result in stationary jet formation (e.g., \cite{Gammie2003,DeVilliers2005,McKinney2006,Porth2019}). In general, different poloidal magnetic field configurations result in different magnetic flux accretions onto the black hole horizon and jet launching properties (e.g., \cite{Beckwith2008,Beckwith2009,Barkov2011,Narayan2012,McKinney2012,White2020}). In particular, the configuration of multiple poloidal magnetic loops with different polarities leads to intermittent, non-stationary jet formation~(e.g.,~\cite{Nathanail2020,Nathanail2021}). In same-polarity multiple poloidal magnetic loop cases, same-polarity loops quickly reconnect to form a large loop with a resulting flow similar to the single magnetic loop case. On the other hand, opposite-polarity loops preserve their small coherence length. This generates plasmoids and plasmoid chains by magnetic reconnection, which results in flaring activity. Similar plasmoid formation has even been seen in single poloidal loop cases at the jet funnel wall region in general relativistic resistive MHD simulations~\cite{Ripperda2020}. If GRMHD simulations with multiple loops enter a MAD regime, plasmoid chains develop on the equatorial plane near the horizon and a quasi-striped jet is produced~\cite{Chashkina2021}. In long-term evolution, jets become inactive and the accretion flow state transitions to the SANE regime. In general, GRMHD simulations with multiple magnetic loops need higher numerical resolution to resolve plasmoid formation and the development of plasmoid chains. At present, most of these simulations are performed in 2D. Therefore, it is important to investigate how plasmoids and intermittent jets develop in 3D GRMHD~simulations. 

In the general consensus of jet formation via MHD processes, one requires poloidal magnetic fields, hence many GRMHD simulations are performed with poloidal magnetic fields in order to produce relativistic jets. Large-scale poloidal magnetic fields can be formed in situ through a turbulent dynamo \cite{Brandenburg1995} produced by the MRI. In earlier studies, \citet{Beckwith2008} presented jets that cannot be formed from a purely toroidal magnetic field initial condition. In larger torus configurations, \citet{McKinney2012} found that weak jets can form from toroidal magnetic field initial configurations. Recently, \mbox{\citet{Liska2020}} showed that a simulation with a toroidal magnetic field can generate poloidal fields self-consistently via an $\alpha$ dynamo and produce relativistic jets. From these simulations, it has been learned that numerical resolution is important. High numerical resolutions are needed to capture turbulent dynamo processes seen in local shearing box simulations. 

Toroidal magnetic field configurations in geometrically thick torii are used to investigate the stability of the system. \citet{Komissarov2006} described a new torus solution with purely toroidal magnetic fields. If this torus solution is strongly magnetized, the system is unstable to non-axisymmetric MRI \cite{Wielgus2015} and over a few tens of orbital periods, the magnetization of the disk significantly drops before reaching its steady-state value in the weakly magnetized disk \cite{Fragile2017}. \citet{Bugli2018} have studied the effect of the Papaloizou--Pringle instability for a thick torus with a toroidal magnetic field, finding that weak toroidal magnetic fields suppress the development of the Papaloizou--Pringle instability.   

\subsection{Jet Formation from Tilted Thick Torii}

\begin{figure}[H]	
\includegraphics[width=0.59\textwidth]{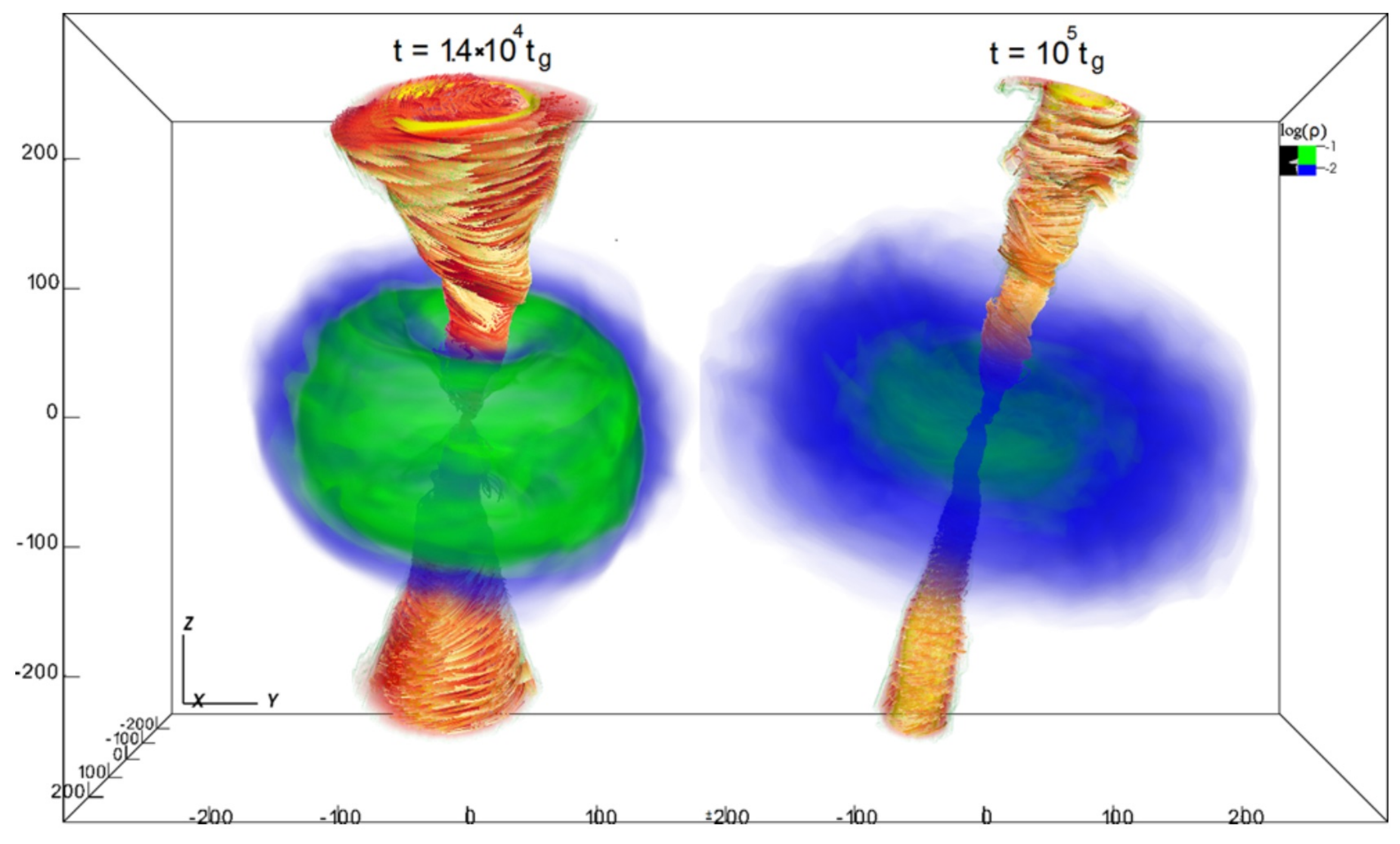}
\caption{3D volume rendering of density (green and blue) and jet magnetic field colored by magnetic energy (red and yellow) in GRMHD simulations of a tilted torus. Figure is reproduced with permission from Liska et al. MNRAS, 474, L81 (2018) \cite{Liska2018}.}
\label{fig:MAD_jet2}
\end{figure}

As discussed in the previous section, misalignment (tilt) between the accretion disk and black hole spin axis brings in another free parameter for jet formation simulations of magnetized torii. The first GRMHD simulations with a tilted torus were performed by \citet{Fragile2007}. Tilted disk configurations present more complicated dynamics than untilted systems. In such simulations, the main body of the disk remains tilted and there is no indication of a Bardeen--Petterson effect in the disk at large. When the tilt angle becomes large, the tilted disk can develop standing shocks that can facilitate the transport of angular momentum and dissipation of energy in the disk \cite{Fragile2008}. Recently, \mbox{\citet{White2019b}} performed a parameter survey of tilted torii with different black hole spin parameters and inclination angles. In their simulations, magnetized polar outflows form along the disk rotation axis, in agreement with \citet{Liska2018} for tilted thin disk simulations (see Fig.~\ref{fig:MAD_jet2}). Similar results are also obtained by \citet{Chatterjee2020}. A parameter study by \citet{White2019b} confirmed the presence of standing shocks at large inclination angles and no observable Bardeen--Petterson effect. Many GRMHD simulations suggest that tilted torus simulations can produce relativistic jets as seen in non-tilted torus simulations. However, the jet propagation direction is still not fully understood. The jet propagation direction will affect the interpretation of black hole shadow images, since the inferred observer inclination angle depends on the jet orientation in the sky. 

%

\subsection{Geometrically Thick Torus Simulations in Non-Ideal GRMHD}

In RIAFs, the Coulomb mean free paths of both ions and electrons are much larger than the typical length scale of disks, $\sim$$r_{\rm g}$ \cite{Mahadevan1997}. Therefore, the plasma in the disks is expected to be collisionless. This raises questions about the validity of the ideal MHD approximation in RIAFs. One extension of ideal MHD approaches is the so-called weakly collisional plasma model. In this approach, non-ideal effects are treated as perturbations relative to an ideal fluid. This includes anisotropic heat and momentum transport. \mbox{\citet{Chandra2015}} have described a covariant form for a weakly collisional magnetized plasma and developed a new extended GRMHD code \cite{Chandra2017}. Application to the accretion torus in RIAFs has been performed by {\citet{Foucart2016,Foucart2017}}. They found that pressure anisotropy produces outward angular momentum transport with a magnitude comparable to that of MHD turbulence in the disk, along with a significant increase in temperature at the funnel wall region. They also found that the heat flux is dynamically unimportant. These simulations in extended GRMHD have shown similarities to those in ideal GRMHD simulations. Therefore, accretion flows neglecting non-ideal effects are likely reasonable even if the accretion flows are nearly collisionless.  

Another extension from ideal MHD is adding resistivity. In ideal GRMHD simulations of magnetized accretion flows, dissipation is seen at the grid scale due to numerical resistivity (e.g., \cite{Nathanail2020}). To study magnetic reconnection and plasmoid formation, it is therefore better to consider reconnection physically. Recently, {\citet{Ripperda2019}} developed a resistive GRMHD simulation code. In that simulation, an explicit finite resistivity acts as a proxy for kinetic effects, presenting a physical model of magnetic reconnection and plasmoid formation in turbulent black hole accretion flows \cite{Ripperda2020}. 
{\citet{Tomei2020} have investigated a mean-field dynamo in resistive GRMHD simulations of a magnetized accretion torus in the full non-linear regime. The dynamo process produces an exponential growth of initial seed magnetic field \cite{Bugli2014}. Different dynamo coefficients provide different growth rates, although the magnetic field amplification seems to saturate at similar levels.}
Resistive GRMHD simulations have similar difficulties to radiation GRMHD, because the resistive term provides a stiffness in the equations. Therefore, the stiff term needs to be solved~implicitly.   

\section{Jet Formation in Radiative GRMHD Simulations}\label{sec4}

GRMHD simulations of jet launching have been performed from magnetized thick accretion torii for RIAFs. In accretion theory, depending on the mass accretion rate, radiation effects become important near the Eddington limit. When considering accretion flows with mass accretion rates close to the Eddington limit, a proper treatment of radiation is~crucial.  

One rather simple treatment of radiation is through considering radiative cooling. A simple local cooling prescription via different radiation processes is implemented in GRMHD codes (e.g., \cite{Fragile2009,Dibi2012,Yoon2020}). \citet{Dibi2012} have identified that even for mass accretion rates of $\dot{M}/\dot{M}_{\rm Edd} \approx 10^{-7}$, radiative losses may play an important role in GRMHD simulations, where $\dot{M}_{\rm Edd}$ is the Eddington mass accretion rate. 

Another approach for radiative cooling is using a full frequency- and angle-dependent Monte Carlo treatment of the radiation field \cite{Ryan2015,Ryan2017,Ryan2018,Yao2021}, coupled with GRMHD simulations. These have been applied to simulations of RIAFs, obtaining similar results and concluding that global radiative effects play a sub-dominant yet non-negligible role in disk dynamics if $\dot{M}/ \dot{M}_{\rm Edd} \ge 10^{-6}$, as suggested by \citet{Dibi2012}.

\begin{figure}[H]	
\includegraphics[width=0.69\textwidth]{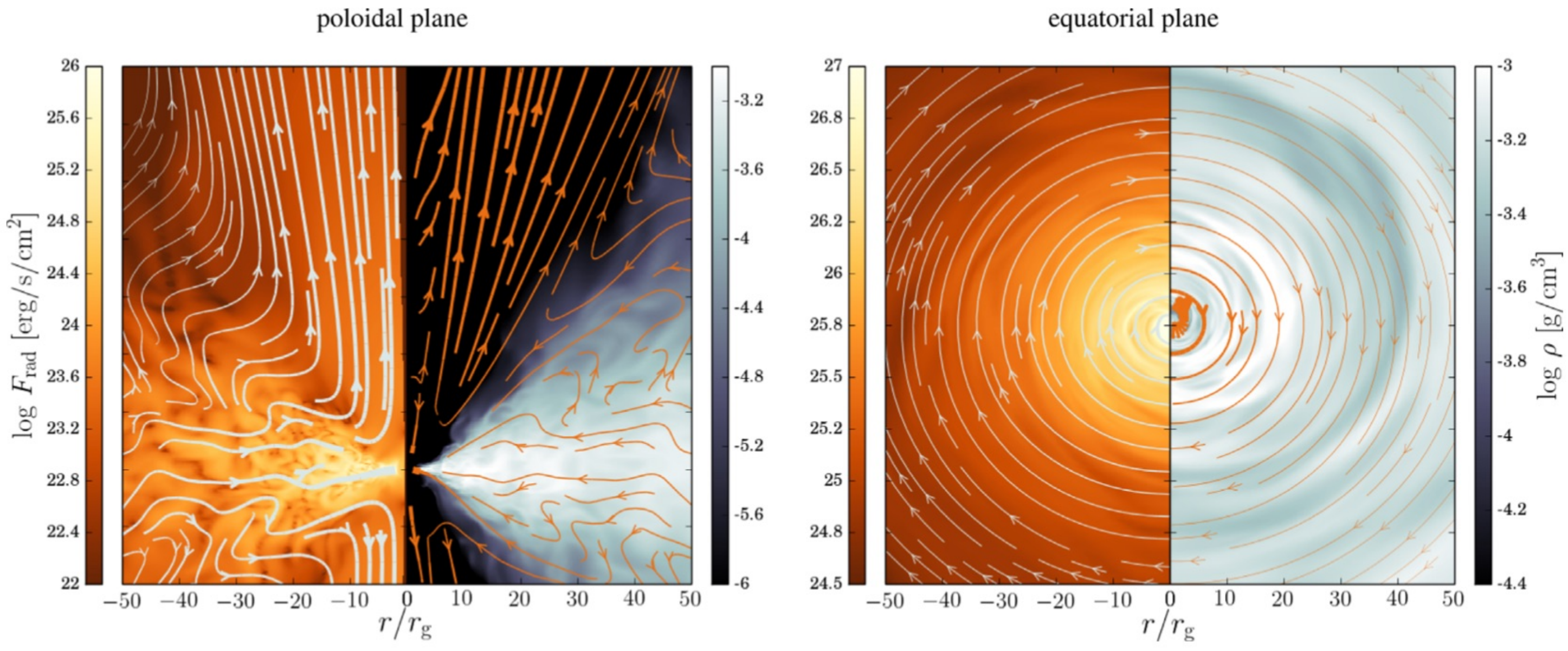}
\caption{Snapshot images of radiative flux (left half of each panel) and density (right half of each panel) in the radiation GRMHD simulations of a thick torus. The left and right panels correspond to vertical and equatorial slices, respectively. Lines indicate azimuth- and time-averaged radiative flux and gas velocity. Figure is reproduced with permission from Sadowski and Narayan, MNRAS, 456, 3929 (2016) \cite{Sadowski2016}.} \label{fig:radiation_jet}
\end{figure}

Full coupling with radiation in GRMHD simulations is challenging due to the stiffness of the radiative term. \citet{Farris2008} developed a formalism for incorporating radiation in the Eddington approximation (flux-limited diffusion approximation). Such formalisms have been implemented in several GRMHD codes \cite{Fragile2012}. However, the Eddington approximation cannot handle optically thin flows accurately. More advanced methods have since been developed (e.g., \cite{Sadowski2013,McKinney2014}). These methods solve the radiation momentum equations using an M1 closure scheme. They perform GRRMHD (the second “R” denoting “radiative”) simulations of super-critical accretion disks (e.g., \cite{Sadowski2014,Sadowski2015,Sadowski2016,McKinney2015}).  
These simulations have shown that even super-Eddington accretion disks around spinning black holes produce low-density funnels, with large fractions of energy being extracted from the black hole's rotational energy through a process similar to the BZ mechanism (see Figure~\ref{fig:radiation_jet}). Importantly, these simulated systems have a high radiative efficiency which significantly exceeds the efficiency predicted by slim disk models for these mass accretion rates. The magnetized jet creates a low-density channel for radiation and pushes away the more opaque wind. This leads to radiation flux escaping from the disk and reduces the conversion of radiation energy flux into kinetic energy flux of the wind. Such a mechanism enables high radiative efficiencies in supper-Eddington accretion flows. 

%

\section{Jet Formation without Disk}\label{sec5}

In general setups for the simulation of jet formation, there is a hydrostatic equilibrium of the thin disk or thick torus with the poloidal magnetic field surrounding the central black hole (e.g., \cite{Porth2019}). One general question which is raised is whether an accretion disk or a torus is required for jet formation. If the accreting matter has angular momentum, matter piles up on the equatorial plane and creates a disk-like structure. If this accreting matter involves poloidal magnetic fields, jet-like outflows develop from the resulting disk-like structure via ordinary MHD processes (e.g., \cite{Mizuno2004a,Mizuno2004b}). \citet{Garain2020} performed GRMHD simulations of poloidal magnetic fields advected within low angular momentum accretion flows, yielding outflows from the centrifugal barrier near the horizon. Even GRMHD simulations of spherical accretion flows with magnetic fields have been investigated \cite{Ressler2021}. Though the advection of uniform, relatively weak magnetic fields and a spinning black hole, magnetic reconnection-driven turbulence develops on the equatorial plane, nearly reaching the MAD state and forming electromagnetic jets in the polar region. A jet is formed even when there is no initial net vertical magnetic flux, since turbulent, horizon-scale fluctuations can generate a net vertical magnetic field locally. 

\citet{Ressler2020b} considered a more realistic situation in Sgr A* and performed GRMHD simulations of accretion flows fed by $\sim$30 Wolf--Rayet (WR) stellar winds. The initial conditions of these simulations were provided by larger-scale MHD simulations~\cite{Ressler2020a}. WR stellar winds provide weak magnetic field at large scales. These are amplified by flux freezing and compression within the inflowing gas before reaching the horizon. The accretion flow enters the MAD state through continuous accretion of coherent magnetic fields. The amplified magnetic field then helps to drive polar outflows. 

Several GRMHD simulations have shown that accretion disks or torii are not initially required for jet formation. The jets which form in these simulations are not persistent in some cases. An outstanding issue is the investigation of the similarities and differences between accretion and outflow dynamics in different physical conditions.
This will help establish the critical requirements for jet formation.

\section{Jet Modeling from GRMHD Simulations}\label{sec6}

When matter accretes onto a black hole, it heats up and begins to radiate. Since the radiation within the black hole photon orbit falls into the horizon and never reaches us, the presence of the horizon casts a “shadow”. In GR, the size of this shadow only depends on the mass and spin of the black hole. Therefore, the direct observation of a black hole is a very promising approach to investigate the properties of a black hole and its surrounding plasma dynamics, including the accretion flow and the formation of jets (e.g., \cite{Goddi2017,Younsi2016,Bronzwaer2021}).

General relativistic radiative transfer (GRRT) calculations coupled with the calculation of geodesics in the black hole spacetime are an essential tool for determining the images, spectra, and light curves from matter in the vicinity of BHs.
Without scattering, the integration of the radiation transfer equation can be performed by dividing each ray into a series of small steps. This is the so-called ray-tracing method. Several GRRT codes have been developed to utilize this ray-tracing method (e.g., \cite{Fuerst2004,Noble2007,Dexter2009,Younsi2012,Chan2013,Dexter2016,Chan2018,Pu2016,Bronzwaer2018,Moscibrozdka2018,Bronzwaer2020,Kawashima2021}). Some of these GRRT codes are coupled with GRMHD simulation codes to produce observables such as images, spectra, light curves, and polarization.  


The horizon scale image of a black hole accretion system is closely related to the origin of jets and their structure.
The horizon scale image at mm/sub-mm wavelengths for M87 was first discussed in \citet{Broderick2009}. 
In order to model the jet structure, \citet{Broderick2009} assumed a force-free jet model and plasma loading at a certain height above the horizon. A modified version of the jet model, taking into account the jet terminal Lorentz factor, was considered in \citet{Lu2014}. How different background jet velocities result in different shearing hot spot features within the jet and disk wind is discussed in \citet{Jeter2020}. 
\citet{Takahashi2018} investigated the large-scale jet image feature for different plasma loadings at the foot point of the jet. In a GRMHD flow, there is a region close to the black hole inside of which matter is accreted and outside of which the matter is accelerated outward. This is the so-called stagnation surface, which is defined by a vanishing poloidal velocity. Such a region is theoretically predicted to be highly magnetized and exhibit low densities, providing an ideal site for energetic, non-thermal, and non-ideal processes. Provided that energetic electrons could be injected from the stagnation surface, \citet{Pu2017} have considered the subsequent plasma cooling coupled with a GRMHD background velocity, and proposed that the emission from this stagnation surface could be an observable feature if the jet launching mechanism originates from GRMHD processes.


Calculating the emission from accreting matter requires predicting the electron distribution function. In a thermal synchrotron radiation model, a Maxwell--J\"{u}ttner distribution function with a particular electron temperature is assumed. In an RIAF, asymmetric heating and inefficient Coulomb coupling between electrons and protons lead to a decoupling of electron and proton temperatures. Most GRMHD simulations only consider a single fluid, which essentially describes the ions (mass, temperature, and energy). Therefore, the emitting electron temperature is not constrained. In order to overcome this issue, typically, the electron-to-ion temperature ratio is prescribed manually in post-processing GRRT calculations of GRMHD simulations (e.g., \cite{Shcherbakov2012,Dexter2012,Moscibrodzka2014,Moscibrodzka2016,Chan2015,Gold2017}). The standard prescription for the electron temperature assumes a constant fraction of the ion temperature (e.g., \cite{Moscibrodzka2013,Moscibrodzka2014,Chan2015}). In many cases, the emitting region can be broken down into components according to different physical properties, such as the disk and the jet/funnel. \citet{Moscibrodzka2016} proposed a parametrized electron-to-ion temperature ratio formula which follows the plasma beta distribution of GRMHD simulations. This is the so-called R-$\beta$ prescription. This temperature prescription leads to a hotter electron temperature within more magnetized regions, i.e., jet regions, helping to produce the observed flat radio spectra and making the jet more prominent compared to the accretion disk. Other parametrized prescriptions for the electron temperature are proposed in other studies \cite{Sharma2007,Shcherbakov2012,Gold2017,Anantua2020} which are based on energy balance arguments and the properties at larger radii.
Models inspired by the results of electron thermodynamics in GRMHD simulations have also been considered in recent years (e.g., \cite{Ressler2015,Mizuno2021}). 
The first model of M87 based on GRMHD simulations was presented by \citet{Dexter2012}. Their models included a thermal electron population in the disk and a power-law-based electron distribution in the jet.
\citet{Moscibrodzka2016} reproduced the characteristics of the M87 radio core, namely a flat spectrum and an increasing image size with observing wavelength, by adopting a two-temperature accretion flow with a hot isothermal jet.


\begin{figure}[H]	
\includegraphics[width=0.60\textwidth]{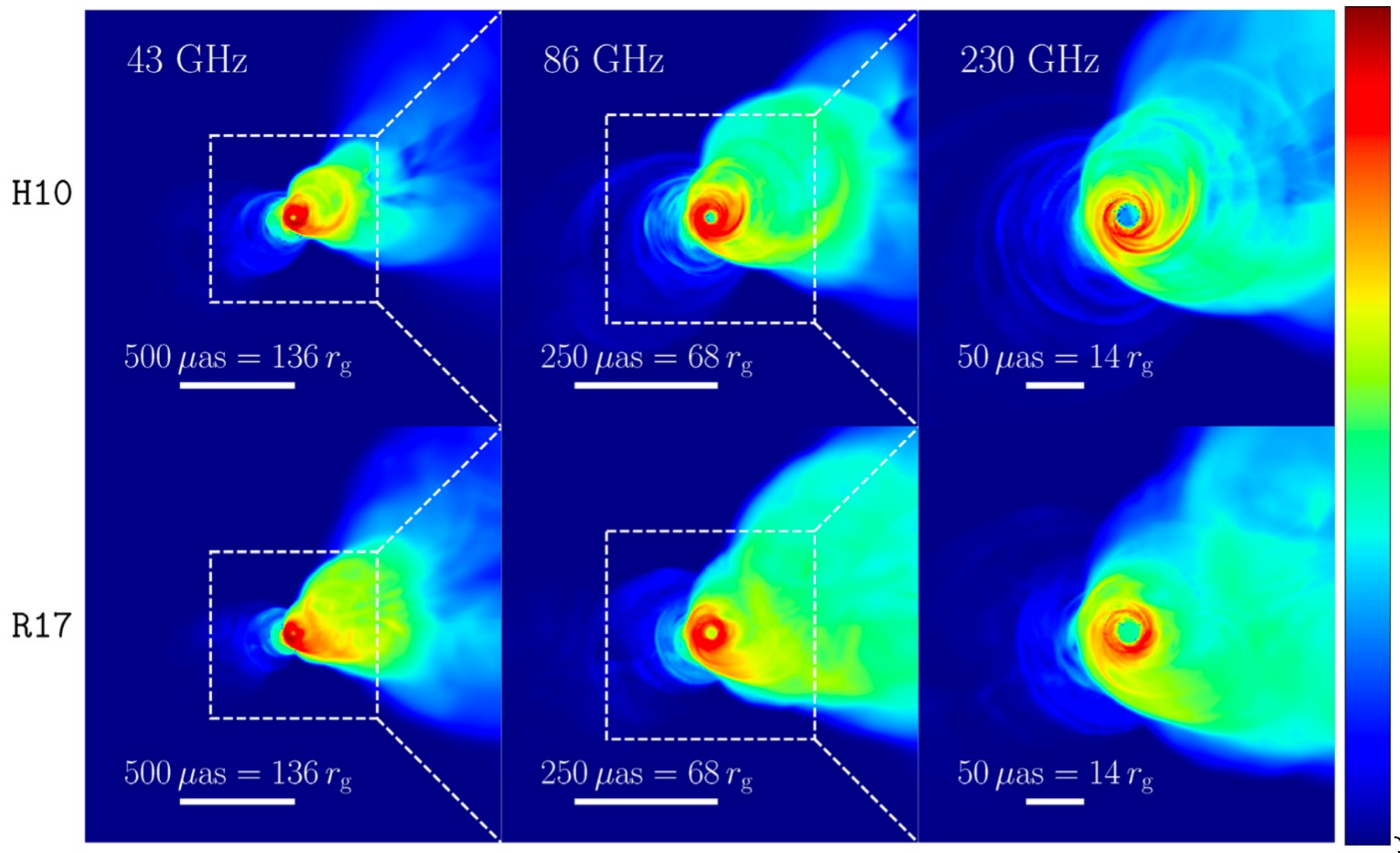}
\caption{Logarithmic scale radiation 
 images of two-temperature GRMHD simulations of a MAD torus with electron heating prescription at 43 \~GHz (\textbf{left}), 86 \~GHz (\textbf{middle}), and 230 \~GHz (\textbf{right}). Snapshots were observed at an inclination angle of 17 degrees with respect to the simulation south pole and rotated 108 degrees counterclockwise, in order to fit the observed jet orientation. 
Figure is reproduced with permission from Chael et al. MNRAS, 486, 2873 (2019) \cite{Chael2019}.} \label{fig:MAD_M87_jet_image}
\end{figure}  
%
\textls[-5]{A more self-consistent approach for obtaining electron temperature from GRMHD simulations involves coupling with electron thermodynamics, where one evolves an electron entropy equation which takes into account local sub-grid electron heating \mbox{\citep{Ressler2015,Ryan2015,Chael2018,Chael2019,Ryan2018,Dexter2020,Mizuno2021}}.} 
In this approach, the back reaction of electron pressure on the dynamics of the accretion flow is neglected (see \cite{Ressler2015}). For the electron heating prescription, this study used two major heating models, turbulent heating (e.g., \cite{Howes2010,Kawazura2019}) and magnetic reconnection heating \mbox{(e.g., \cite{Rowan2017,Rowan2019})}. \citet{Mizuno2021} confirmed that the commonly used parametrized electron-to-ion temperature ratio prescription R-$\beta$ model is well matched to both turbulent and magnetic reconnection electron heating models, when comparing with images at 230\~GHz.
Recently, two-temperature radiation GRMHD simulations have been performed for M87, taking into account the dynamical importance of the photon field on the accretion structure as well as electron cooling \cite{Chael2019,Ryan2018}. In particular, \citet{Chael2019} have found that MAD GRMHD simulations with high BH spin produce wide opening angle jets consistent with VLBI images at 43 and 86\~GHz (see Fig.~\ref{fig:MAD_M87_jet_image}). Similar results are found from MAD GRMHD simulations using a hybrid electron distribution function \cite{Cruz-Osorio2021,Fromm2021}.     


One of the major open questions in modeling the electromagnetic radiation emerging from accretion flows and jets is the shape of the electron distribution function (eDF). The common assumption is that the electrons in the full simulation domain are in a thermal-relativistic Maxwell-J\"{u}ttner distribution. However, this assumption likely breaks down in regions where non-ideal MHD effects are important. 

\citet{Davelaar2018} applied the $\kappa$-eDF in jet regions of axisymmetric 2D GRMHD simulations. The $\kappa$-distribution function is a combination of a relativistic thermal and a relativistic non-thermal power-law distribution, and describes accelerated electrons \mbox{(e.g., \cite{Vasyliunas1968,Xiao2006,Pandya2016,Marszewski2021})}. They found that $\kappa$-jet models increase the radio-emitting region size and radio flux for decreasing values of the $\kappa$ parameter, which corresponds to a larger amount of accelerated electrons in the jet region. In \citet{Davelaar2019}, this work is extended to using 3D GRMHD simulations and a variable $\kappa$ model which is based on sub-grid particle-in-cell (PIC) simulations of trans-relativistic magnetic reconnection \cite{Ball2018}. This result shows that $\kappa$-eDF models reproduce the broad band spectrum from radio to optical wavelengths of M87, which cannot be produced from thermal eDFs. Recently, \citet{Cruz-Osorio2021} investigated the impact of a non-thermal emission models on the structure and morphology of the M87 jet and its spectrum by using MAD GRMHD simulations. They found that MAD models can explain both the observed jet width of M87 at 86 \~GHz (e.g., \cite{Kim2018}) and its~spectrum. 

Current modeling studies based on GRMHD simulations are dominated by RIAFs. Jets in low-luminosity AGNs, in particular M87, have been observed from Mpc scales to sub-pc scales in high structural detail. There is strong motivation to model such jets and extract fundamental physics. Recent GRMHD simulations have been extended to more varied setups, including thin disks and slim disks. GRRT calculations coupled  with these GRMHD simulations will provide jet modeling in a wide variety of AGN jets.

\section{Discussion and Summary}\label{sec7}

Over the past few decades, many GRMHD codes have been developed and applied to study relativistic jet formation in various physical conditions, from geometrically thin disks to geometrically thick torii, with additional physical processes such as radiation feedback. It is understood that poloidal magnetic fields near the black hole horizon are a key ingredient required to produce powerful relativistic jets via MHD processes. Relativistic jets are accelerated around a Lorentz factor of less than 100.
However, for further acceleration, additional magnetic energy dissipation mechanisms are required. 

Recent progress from GRMHD codes using AMR and including different physics such as resistivity and radiation have provided a much wider variety of dynamics of accretion flows onto black holes and corresponding jet formation. This growing diversity in modeling will be important for the fundamental understanding of accretion and jet physics.

From the recent progress of millimeter and sub-millimeter VLBI observations from observatories such as the EHT and GMVA, the collimation and acceleration zones of relativistic jets in several AGNs could be observed. Theoretical modeling of relativistic jets from GRMHD simulations is hence becoming increasingly important. GRMHD simulations of jet formation and subsequent modeling will play a crucial role in the understanding of relativistic jet properties, including jet formation and acceleration mechanisms, which may all be investigated via comparison with future mm- and sub-mm VLBI observations.

\vspace{6pt} 




\funding{This research is supported by the ERC synergy grant ‘BlackHoleCam: Imaging the Event Horizon of Black Holes’ (grant number 610058).}

\dataavailability{The data underlying this article will be shared on reasonable request to the corresponding author.} 

\acknowledgments{I would like to thank Ziri Younsi for useful discussions. This research has made use of NASA’s astrophysics data system (ADS).}

\conflictsofinterest{The authors declare no conflict of interest.} 



\abbreviations{Abbreviations}{
The following abbreviations are used in this manuscript:\\

\noindent 
\begin{tabular}{@{}ll}
ADAF & Advection-Dominated Accretion Flow \\
AGN & Active Galactic Nuclei\\
AMR & Adaptive Mesh Refinement \\
BH & Black Hole\\
EHTC & Event Horizon Telescope Collaboration \\
GR & General Relativity \\
GRMHD & General Relativistic Magnetohydrodynamics \\
GRRT & General Relativistic Radiative Transfer \\
MAD & Magnetically Arrested Disk \\

%

MHD & Magnetohydrodynamics \\
MRI & Magnetorotational Instability \\
SANE & Standard Accretion and Normal Evolution\\
PFD & Poynting Flux Dominated \\
RIAF & Radiatively Inefficient Accretion Flow \\
WR & Wolf--Rayet
\end{tabular}}

%
%

\end{paracol}
\reftitle{References}


\externalbibliography{yes}
\bibliography{ref}

%


\end{document}